\documentstyle[11pt,fleqn]{article}
\title{Comments on "Microscale flow visualization"} 
\author{A. Kwang-Hua Chu \thanks{Present Address : P.O. Box 39, Tou-Di-Ban, Xihong Road,
Urumqi 830000, PR China.  }}  
\date{P.O. Box 30-15, Shanghai 200030, PR China}
\begin{document}           
\maketitle
\begin{abstract}
We make comments on the presentation of Sinton's paper
(Microfluidics and Nanofluidics {\bf 1}: 2, 2004) about the
microscale flow visualization since the effects of the roughness
along the microfabricated wall upon the current macroflow
visualization methods could be significant and cannot be neglected
in microdomain and even nanodomain.
\end{abstract}
\doublerulesep=6mm    %
\baselineskip=6mm
\bibliographystyle{plain}               
\noindent Sinton just presented a rather interesting and
comprehensive review about the microscale flow visualization
(Sinton, 2004) due to advances in microfluidic and nanofluidic
technologies (Li, 2004) being paralleled by advances in methods
for direct optical measurement of transport phenomena on these
scales. As Sinton noticed, a variety of methods for microscale
flow visualization have appeared and evolved since the late 1990s.
These methods and their applications to date are reviewed therein
(Sinton, 2004)  in detail, and in context of the both the
fundamental phenomena they exploit and the fundamental phenomena
they are applied to measure. Where possible, links to macroflow
visualization methods are established, and the physical mechanisms
underlying these methods are explained. \newline We all know that
direct flow visualization is of key importance for the fundamental
understanding of microflows, analyzing, developing and evaluating
novel microfluidic processes, investigating non-ideal behavior
such as spatial and temporal gradients in surface and fluid
properties, and providing benchmark data for computational
investigations (Sinton and Li, 2004). There are, however, some
differences between macro- and microdomain which will influence
the results based on particle-based, scalar-based and
point-detection scanning microfluidic flow visualization methods.
One specific example is the roughness of the bounded wall, which
is quite significant in common microchannels but could be {\it ad
hoc} neglected in macrochannels (Chu, 2000/2002). For
microfabricated Si-based walls, the roughness cannot be eliminated
completely and thus induce quite random scattering effects (e.g.,
for the particle image velocimetry (PIV), the motion of the bulk
fluid is inferred from the observed velocity of marker particles,
Adrian 1991) especially for the particle-based flow visualization
methods. As claimed before (Santiago {\it et al.}, 1998), the
stochastic influence of the Brownian motion of the small particles
was significant, however, ensemble averaging over several images
was shown to greatly improve the obtained velocity field. It is
clear that the near-wall flow field would have poor resolution due
to the irregular scattering coming from the random roughness along
the wall or the confined boundary (Chu, 2000/2002). \newline How
about the scalar-based flow velocimetry, where the motion of the
bulk fluid is inferred from the observed velocity of a conserved
scalar? What is the roughness effect in microchannels for the
scattering of light emitting molecules (fluorescent or
phosphorescent) which are typically employed to increase the
signal from relatively small volumes of fluid (for this same
reason fluorescently labeled particles are employed in
particle-based visualization of microflows). One example is,
analysis methods for obtaining velocity data from the observed
transport of a conserved scalar which are collectively termed
scalar image velocimetry (SIV). The base mechanisms employed in
these scalar-based flow velocimetry techniques are, fluorescence,
photobleached fluorescence, photochromic reaction,
phosphorescence, caged fluorescence, and IR heating. The essential
near-wall flow field will be poorly resolved once the microchannel
geometry is narrowed down compared to the intrinsic random
roughness produced by the current microfabrication technology
(Dwivedi, 2000;  Komvopoulos, 1996). The situations will be worse
considering the gaseous flow visualization in microchannels (as
noted before, owing to challenges associated with seeding and
particle inertia, micro-PIV has not been successfully applied to
gaseous microflows to date; Wereley and Meinhart, 2004). \newline
As commented by Sinton (2004) :  it is expected, however, that
these distinctions will become progressively blurred as more new
and hybrid techniques are developed. Particularly as marker sizes
are reduced to the limit of a single molecule, scalar and particle
distinctions are no longer meaningful. The majority of efforts in
this area have involved liquid flows, however, recent developments
toward gaseous microflow visualization were also discussed.
Selection of an appropriate microscale visualization method
depends chiefly on the phenomena and application of interest.
Micro-PIV is the most well-developed microscale flow visualization
method. The ability to acquire high spatial/temporal resolution
velocity field data in multidimensional microflows has driven this
development. Although micro-PIV technology has matured rapidly and
commercial systems have been available for some time, significant
improvements continue to appear in the literature. A variety of
scalar-based methods for microscale flow visualization have been
developed and applied to study microflows (the most well-developed
of which is caged-fluorescence imaging). Although physical
flow-tagging mechanisms vary, scalar- based methods typically
involve tracking a crossstream flow marker and thus are most
suitably applied to unidirectional microchannel flows. One
advantage of scalar-based methods is that the nature of the
velocity field may be interpreted readily from the image data.
Thus, scalar-based methods are particularly applicable when
species transport is of interest, a common focus in microfluidic
chip applications. Point-detection scanning based microflow
visualization methods have shown promise with respect to spatial
resolution, and optical sectioning capability. Temporal resolution
is typically limited by the scanning rate, however, significant
improvements have been made in this regard recently. {\it Most of
the microscale flow visualization methods discussed have evolved
from methods developed originally for macroscale flows. It is
unlikely, however, that developed microscale flow visualization
methods will be translated to nanoscale flows in a similar manner.
Resolving nanoscale features with visible light presents a
fundamental challenge.} Although point-detection scanning methods
have potential to increase the flow measurement resolution on the
microscale, spatial resolution is ultimately limited by the
optical probe volume (length scale on the order of 100 nm), which,
in turn, is limited by the wavelength of light employed. In that
context, optical-based spatially resolved flow measurements in
nanochannels are, at best, difficult to visualize. Both the future
refinement of microscale flow visualization methods and the
development of direct flow measurement methods for nanoflows will
be followed with great interest. \newline The present author
believes that the limitation of the current micro- and
nanofabrication technology : how to smooth out the random
roughness along the wall or the confined boundary (Zubel and
Kramkowska, 2001)  will also produce challenges to the researchers
working on the point-detection scanning techniques for microflows
and/or nanoflows (Li, 2004).

\end{document}